\documentclass{PoS}

\usepackage{url}
\usepackage[utf8]{inputenc}
\usepackage{mathtools}
\usepackage{slashed}
\usepackage{floatrow}

\newlength{\bibitemsep}
\newlength{\bibparskip}
\let\oldthebibliography\thebibliography
\renewcommand\thebibliography[1]{%
  \oldthebibliography{#1}%
  \setlength{\parskip}{\bibitemsep}%
  \setlength{\itemsep}{\bibparskip}%
}

\def\eq#1{Eq.~(\ref{#1})}
\def\fig#1{Fig.~\ref{#1}}
\def\tbl#1{Table~\ref{#1}}
\def\sec#1{Sec.~\ref{#1}}

\title{Nucleon electromagnetic form factors at high $Q^2$ from Wilson-clover fermions}

\ShortTitle{Nucleon EM form factors at high $Q^2$}

\author{\speaker{Christos Kallidonis}\\
       Department of Physics and Astronomy, Stony Brook University, Stony Brook, NY 11794, USA\\
       E-mail: \email{christos.kallidonis@stonybrook.edu}}

\author{Sergey Syritsyn\\
        Department of Physics and Astronomy, Stony Brook University, Stony Brook, NY 11794, USA, \\
        RIKEN BNL Research Center, Brookhaven National Laboratory, Upton, NY 11973, USA\\
        E-mail: \email{sergey.syritsyn@stonybrook.edu}}

\author{Michael Engelhardt\\
        Department of Physics, New Mexico State University, Las Cruces, NM 88003-8001, USA \\
        E-mail: \email{engel@nmsu.edu}}
        
\author{Jeremy Green\\
        NIC, Deutsches Elektronen-Synchrotron, D-15738 Zeuthen, Germany\\
        E-mail: \email{jeremy.green@desy.de}}

\author{Stefan Meinel\\
       Department of Physics, University of Arizona, Tucson, AZ 85721, USA,\\
       RIKEN BNL Research Center, Brookhaven National Laboratory, Upton, NY 11973, USA\\
        E-mail: \email{smeinel@email.arizona.edu}}

\author{John Negele, Andrew Pochinsky\\
        Center for Theoretical Physics, Massachusetts Institute of Technology, Cambridge, MA 02139, USA\\
        E-mail: \email{negele@mit.edu,avp@mit.edu}}

\abstract{We present results on the nucleon electromagnetic form factors from Lattice QCD at
momentum transfer up to about $12$~GeV$^2$. We analyze two gauge ensembles with the
Wilson-clover fermion action, a lattice spacing of $a\approx 0.09$~fm and pion masses
$m_\pi\approx 170$~MeV and $m_\pi\approx 280$~MeV. In our analysis we employ momentum smearing as well as a set of techniques to investigate excited state effects. Good agreement with experiment and phenomenology is found for the ratios $G_E/G_M$ and $F_2/F_1$, whereas discrepancies are observed for the individual form factors $F_1$ and $F_2$. We discuss various systematics that may affect our calculation.}

\FullConference{The 36th Annual International Symposium on Lattice Field Theory - LATTICE2018\\
		22-28 July, 2018\\
		Michigan State University, East Lansing, Michigan, USA.}

\begin{document}

\section{Introduction}

The electric and magnetic form factors of the nucleon are important probes of its internal structure, as they are intimately related to its distributions of electric charge and magnetization. They have been precisely measured from elastic electron-proton scattering experiments since the 1950s. More recent experiments include the ones at JLab, MIT-Bates and Mainz. For a recent review on electron elastic scattering experiments, see Ref.~\cite{Punjabi:2015bba}. 
Concerning the region $Q^2 \gg m_N^2$, experimental data of the $G_E/G_M$ and $F_2/F_1$ ratios currently exist up to $Q^2\approx 8.5$~GeV$^2$ for the proton and up to $Q^2\approx 3.4$~GeV$^2$ for the neutron. The extensive physics program underway at the upgraded Continuous Electron Beam Accelerator Facility (CEBAF) in JLab will explore proton and neutron form factors up to $Q^2=18$~GeV$^2$, see e.g. Refs.~\cite{deJager:2009xs,Riordan:2010zz}. Due to the ongoing and planned experimental activity, a lattice QCD calculation of the electromagnetic (EM) form factors at high $Q^2$ is particularly timely.
Moreover, an \emph{ab initio} calculation can test the validity of various phenomenological and perturbative QCD (pQCD) calculations that predict the form factors' $Q^2$-dependence beyond the region currently available from experiments, as well as provide crucial input to Deeply Virtual Compton Scattering (DVCS) experiments, seeking to determine nucleon observables related to Generalized Parton Distributions (GPDs).

We analyze two gauge ensembles with $N_f=2+1$ quark flavors, featuring the Wilson-clover fermion action and the tree-level tadpole-improved Symanzik gauge action. The gauge configurations are generated by the JLab/W\&M collaboration~\cite{Edwards:2016aaa}. Both ensembles have $\beta=6.3$ and $C_{\rm SW} = 1.205366$. One iteration of Stout smearing is performed in the fermion action using $\rho=0.125$ for the weight parameter. We collect the rest of the parameters entering our calculation and the accumulated statistics for each ensemble in~\tbl{tbl:params}. The renormalization constants $Z_V$ of the vector current, taken from Ref.~\cite{Yoon:2016jzj}, are calculated in the non-perturbative RI-sMOM scheme~\cite{Martinelli:1994ty,Sturm:2009kb} at a scale of 2~GeV~\cite{Bhattacharya:2013ehc}. The lattice spacing values were obtained from the Wilson-flow scale $w_0$~\cite{Borsanyi:2012zs}.
\vspace{-0.1cm}
\begin{table}[h]
\begin{center}
\renewcommand{\arraystretch}{0.9}
\renewcommand{\tabcolsep}{5.5pt}
\begin{tabular}{c|cccc|cc|cc|c}
\hline\hline
Ens. & $a$ (fm) & $am_l$ & $am_s$ & $m_\pi$ (MeV) & $L^3\times T$ & $m_\pi L$ & $N_{\rm cfg}$ & Stat. & $Z_V$ \\
\hline
D5 & 0.094(1)  &  -0.2390  &  -0.2050  &  278(3)  & $32^3\times 64$  &  4.2  &  1346  & 86144 & 0.832(8) \\
D6 & 0.091(1)  &  -0.2416  &  -0.2050  &  166(2)  & $48^3\times 96$  &  3.7  &  784   & 50176 & 0.826(9) \\
\hline \hline
\end{tabular}
\caption{Input parameters of our lattice simulations with the corresponding statistics accumulated.}
\label{tbl:params}
\end{center}
\vspace*{-0.5cm}
\end{table} 

\vspace{-0.1cm}


\section{Form factor extraction}

The electromagnetic form factors are extracted from the matrix element of the vector current
\begin{equation}\label{eq:ff_decomp}
\langle N(p',s')|\mathcal{O}_\mu^V|N(p,s)\rangle = \bar{u}(p',s')\left[\gamma_\mu F_1(q^2) + \frac{i\sigma_{\mu\nu}q^\nu}{2m_N}F_2(q^2)\right]u_N(p,s)\;,
\end{equation}
where $N(p,s)$ is the nucleon state with momentum $p$ and spin $s$, $m_N$ is the nucleon mass, $q=p'-p$ is the momentum transfer from initial ($p$) to final ($p'$) momentum, $u_N$ is the nucleon spinor and $F_1$, $F_2$ are the (elastic) Dirac and Pauli form factors, respectively. We use the local vector current, $\mathcal{O}_\mu^V(x)=\bar{\psi}(x)\gamma_\mu\psi(x)$. The electric $G_E$ and magnetic $G_M$ Sachs form factors are expressed in terms of $F_1$ and $F_2$ as $G_E(q^2) = F_1(q^2) + \frac{q^2}{(2m_N)^2}F_2(q^2)$ and $G_M(q^2) = F_1(q^2) + F_2(q^2)$.

On the lattice, after Wick-rotating to Euclidean time, the matrix element of~\eq{eq:ff_decomp} is extracted from three- and two-point functions, given by
\begin{eqnarray}\label{eq:2pt3pt}
G_\mu(\Gamma;\vec{p}',\vec{q};t_s,t_{\rm ins}) &=& \sum_{\vec{x}_s,\vec{x}_{\rm ins}} e^{-i\vec{p}'\cdot(\vec{x}_s-\vec{x}_0)} e^{i\vec{q}\cdot(\vec{x}_{\rm ins}-\vec{x}_0)}
 \Gamma_{\beta\alpha}\langle J_\alpha (\vec{x}_s,t_s)\mathcal{O}_\mu^V(\vec{x}_{\rm ins},t_{\rm ins})\bar{J}_\beta(\vec{x}_0,t_0)\rangle\;{\rm and} \nonumber\\
C(\vec{p}';t_s) &=& \sum_{\vec{x}_s} e^{-i\vec{p}'\cdot(\vec{x}_s-\vec{x}_0)} 
 (\Gamma_4)_{\beta\alpha}\langle J_\alpha (\vec{x}_s,t_s)\bar{J}_\beta(\vec{x}_0,t_0)\rangle\;,
\end{eqnarray}
respectively. With $\Gamma_4$ we denote the standard unpolarized parity projector, $\Gamma_4 \equiv (1+\gamma_4)/4$, acting on the Dirac indices $\alpha$ and $\beta$. The three-point function is projected with $\Gamma=\Gamma_4$ as well as with the polarized projectors $\Gamma=\Gamma_k \equiv i\gamma_5\gamma_k\Gamma_4$.  We use the standard interpolating field for the proton, $J_\alpha(\vec{x},t) = \epsilon^{abc}u_\alpha^a(x)[(u^b)^\top(x)(C\gamma_5)d^c(x)]$, where $C=\gamma_0\gamma_2$ is the charge conjugation matrix. In order to access the high-$Q^2$ region while keeping the energy of the states as low as possible, we set the sink momentum $\vec{p}'$ to a fixed nonzero value. For each value of momentum transfer $\vec{q}$ the initial momentum is then fixed accordingly as $\vec{p} = \vec{p}' - \vec{q}$. In our analysis, we obtain results for two values of $\vec{p}'$ for the D5 ensemble, $\vec{p}' = \frac{2\pi}{L}(-4,0,0)$ and $\vec{p}' = \frac{2\pi}{L}(-3,-3,0)$, which correspond to $Q^2 = 10.9$~GeV$^2$ and $Q^2 = 12.2$~GeV$^2$ in the Breit frame, respectively. For the D6 ensemble we use $\vec{p}' = \frac{2\pi}{L}(-5,0,0)$, corresponding to $Q^2=8.1$~GeV$^2$ in the Breit frame. We use Gaussian ``momentum'' smeared point sources, as described in Ref.~\cite{Bali:2016lva}, in order to achieve increased signal-to-noise ratio in the boosted frame. In all cases, we set $(N_G,\alpha_G) = (50,2.0)$, and choose $\vec{k}_b = 0.5 \vec{p}'$ for the boost momentum vector in the Gaussian smearing function. When smearing the propagators entering the three-point function at the sink, we use $\vec{k}_b = -0.5 \vec{p}'$. We also apply APE smearing to the gauge links that enter the smearing operator, with parameters $(N_{\rm APE},\alpha_{\rm APE}) = (25,2.5)$. At this point, we stress that throughout our calculation we consider only connected contributions to the three-point functions coming from the up- or down-quark vector current.

In order to increase the statistical precision, we employ the all-mode-averaging (AMA) technique~\cite{Bali:2009hu,Blum:2012uh}. For both the D5 and D6 ensembles we invert $N_{\rm LP}=64$ low-precision sources per configuration, setting the relaxed solver tolerance to $10^{-4}$, and combine them with $N_{\rm HP} = 4$ high-precision sources (inverted to $10^{-10}$) to correct for the bias. We also employ the coherent sequential source method~\cite{Bratt:2010jn}, using two coherent sources for D5 and four for D6. For both D5 and D6, we calculate the three-point function for five values of the source-sink time separation, $t_{\rm sep}\equiv t_s-t_0$, ranging from $t_{\rm sep}\simeq 0.55$~fm to $t_{\rm sep}\simeq 0.95$~fm, in order to examine excited state effects.

Having calculated the two- and three-point functions, we form the following optimized ratio
\begin{equation}\label{eq:ratio2pt3pt}
R_\mu(\Gamma;\vec{p}',\vec{q};t_s,t_{\rm ins}) = \frac{G_\mu(\Gamma;\vec{p}',\vec{q};t_s,t_{\rm ins})}{C(\vec{p}';t_s-t_0)}\times \sqrt{\frac{C(\vec{p};t_s-t_{\rm ins})C(\vec{p}';t_{\rm ins}-t_0)C(\vec{p}';t_s-t_0)} {C(\vec{p}';t_s-t_{\rm ins})C(\vec{p};t_{\rm ins}-t_0)C(\vec{p};t_s-t_0)}}\;.
\end{equation}

\begin{figure}[!ht]
\floatbox[{\capbeside\thisfloatsetup{capbesideposition={right,top},capbesidewidth=0.47\textwidth}}]{figure}[\FBwidth]
{\includegraphics[width=0.48\textwidth]{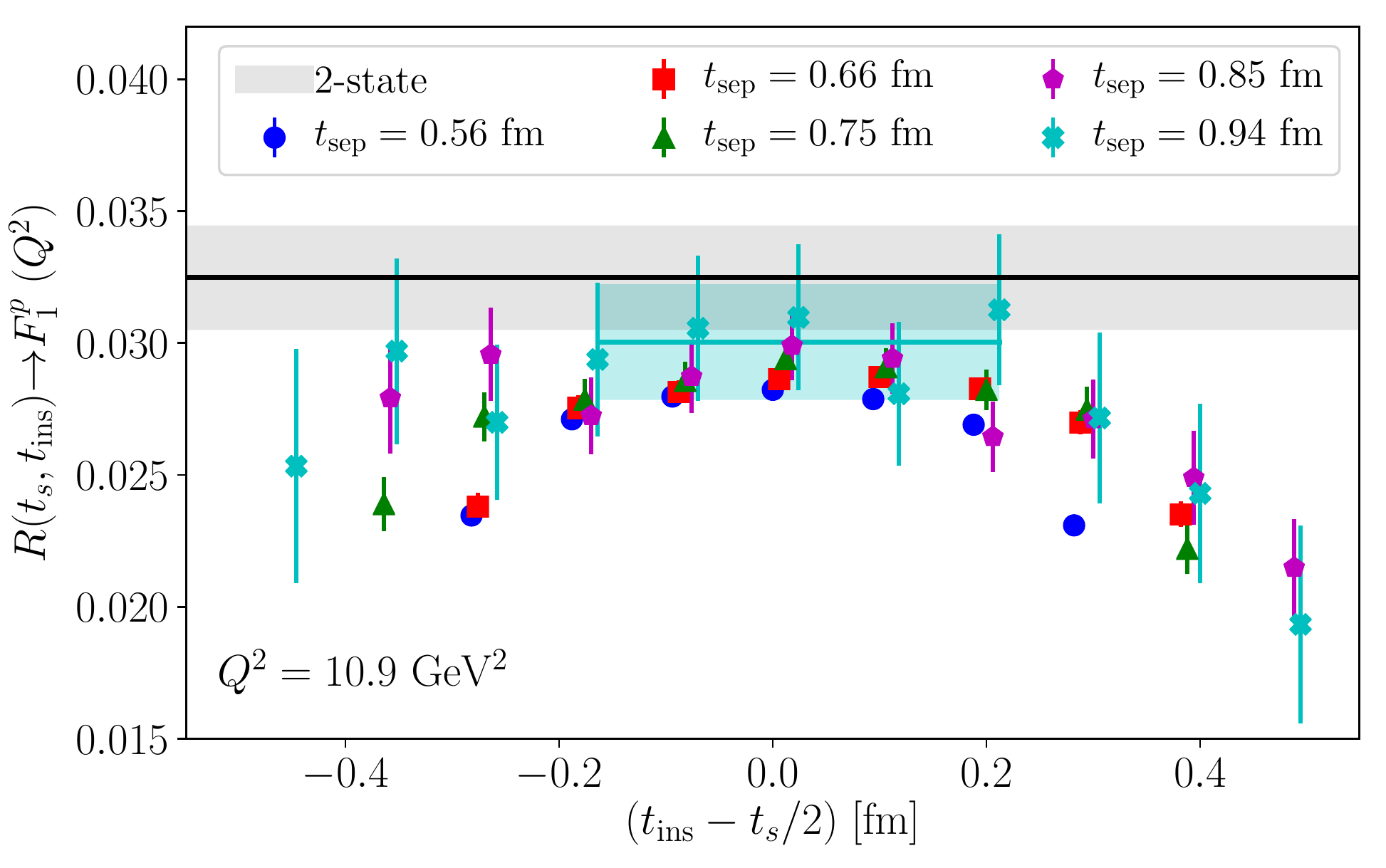}}
{\caption{Ratio yielding the proton $F_1$ form factor at $Q^2=10.9$~GeV$^2$, from our D5 ensemble using sink momentum $\vec{p}'=\frac{2\pi}{L}(-4,0,0)$.}
\label{fig:plateaus}}
\end{figure}

We employ two methods to isolate the desired ground state matrix element. In the so-called \emph{plateau method}, we take the limits $t_s-t_{\rm ins}\gg 1$ and $t_{\rm ins}-t_0\gg 1$ such that the unknown overlaps and exponential factors in~\eq{eq:ratio2pt3pt} cancel out, i.e. $R_\mu(\Gamma;\vec{p}',\vec{q};t_s,t_{\rm ins})\rightarrow \Pi_\mu(\Gamma;\vec{p}',\vec{q})$,
where $\Pi_\mu$ is a time-independent quantity. For each $t_{\rm sep}$ we look for a window of $t_{\rm ins}$ values where the ratio $R_\mu$ forms a plateau. We then fit $R_\mu$ to a constant within the plateau region to obtain $\Pi_\mu$. We consider that excited states are sufficiently suppressed when our fitted value does not change with $t_{\rm sep}$; the matrix element is then the value of $\Pi_\mu$ at the selected $t_{\rm sep}$. This is demonstrated in~\fig{fig:plateaus}, where we plot the ratio $R_\mu$ yielding the proton $F_1$ for our D5 ensemble at $\vec{p}'=\frac{2\pi}{L}(-4,0,0)$ and $Q^2=10.9$~GeV$^2$. As seen, there is a small trend towards larger values of the ratio with increased $t_{\rm sep}$, however they are all compatible in the plateau region. We choose the rather conservative plateau fit at $t_{\rm sep} = 0.94$~fm, shown in the light blue band. The extent of the band indicates the plateau region considered for the fit.
\vspace*{-0.2cm}
\vspace{-0.4cm}

The second technique we employ is the \emph{two-state fit} method. Here, we truncate the two- and three-point functions in~\eq{eq:2pt3pt}, considering contributions up to the first excited state, to obtain
\begin{eqnarray}
G_\mu(\Gamma;\vec{p}',\vec{q};t_s,t_{\rm ins}) &\simeq& e^{-E_0(\vec{p}')(t_s-t_{\rm ins})} e^{-E_0(\vec{p})(t_{\rm ins}-t_0)} \times \left[A_{00}(\vec{p},\vec{p}') + A_{01}(\vec{p},\vec{p}') e^{-\Delta E_1(\vec{p})(t_{\rm ins}-t_0)} + \right. \nonumber\\
&+& \left. A_{10}(\vec{p},\vec{p}') e^{-\Delta E_1(\vec{p}')(t_s-t_{\rm ins})} + A_{11}(\vec{p},\vec{p}') e^{-\Delta E_1(\vec{p}')(t_s-t_{\rm ins})} e^{-\Delta E_1(\vec{p})(t_{\rm ins}-t_0)} \right] \;, \label{eq:3pt_2st} \\
C(\vec{p}';t_s) &\simeq& e^{-E_0(\vec{p}')(t_s-t_0)}\left[c_0(\vec{p}') + c_1(\vec{p}')e^{-\Delta E_1(\vec{p}')(t_s-t_0)} \right]\;, \label{eq:2pt_2st}
\end{eqnarray}
where $\Delta E_1(\vec{p}) \equiv E_1(\vec{p})-E_0(\vec{p})$ is the energy difference between the first excited state and the ground state of the nucleon at momentum $\vec{p}$. The coefficients are $c_n(\vec{p}') = |\langle N|n,\vec{p}'\rangle|^2 / 2E_n(\vec{p}')$ and $A_{nm}(\vec{p},\vec{p}') = \langle N|n,\vec{p}'\rangle\langle m,\vec{p}|N\rangle \langle n,\vec{p}'|\mathcal{O}_\mu^V|m,\vec{p}\rangle / [2 \sqrt{E_n(\vec{p})E_n(\vec{p}')}]$, with $\langle 0,\vec{p}'|\mathcal{O}_\mu^V|0,\vec{p}\rangle$ the desired ground state matrix element. The method first proceeds by performing a fit to the two-point function data using~\eq{eq:2pt_2st} as an Ansatz, to determine $c_0$, $c_1$, $E_0$ and $\Delta E_1$. The next step is to fit the three-point function according to~\eq{eq:3pt_2st}, combining all $t_{\rm sep}$. In this fit we use as input the values of $E_0$ and $\Delta E_1$ determined from the two-point function fit and treat the coefficients $A_{nm}$ as fit parameters. The desired matrix element is then obtained as $\langle 0,\vec{p}'|\mathcal{O}_\mu^V|0,\vec{p}\rangle = A_{00}(\vec{p},\vec{p}')/\sqrt{c_0(\vec{p})c_0(\vec{p}')}$. In~\fig{fig:plateaus} we show the result of our two-state fit with the gray band, and it is compatible with our plateau fit at $t_{\rm sep}=0.94$~fm.

\vspace{-0.2cm}


\section{Results}\label{sec:results}

\begin{figure}[!ht]\vspace*{-0.2cm}
\begin{minipage}{0.33\textwidth}
\includegraphics[width=\textwidth]{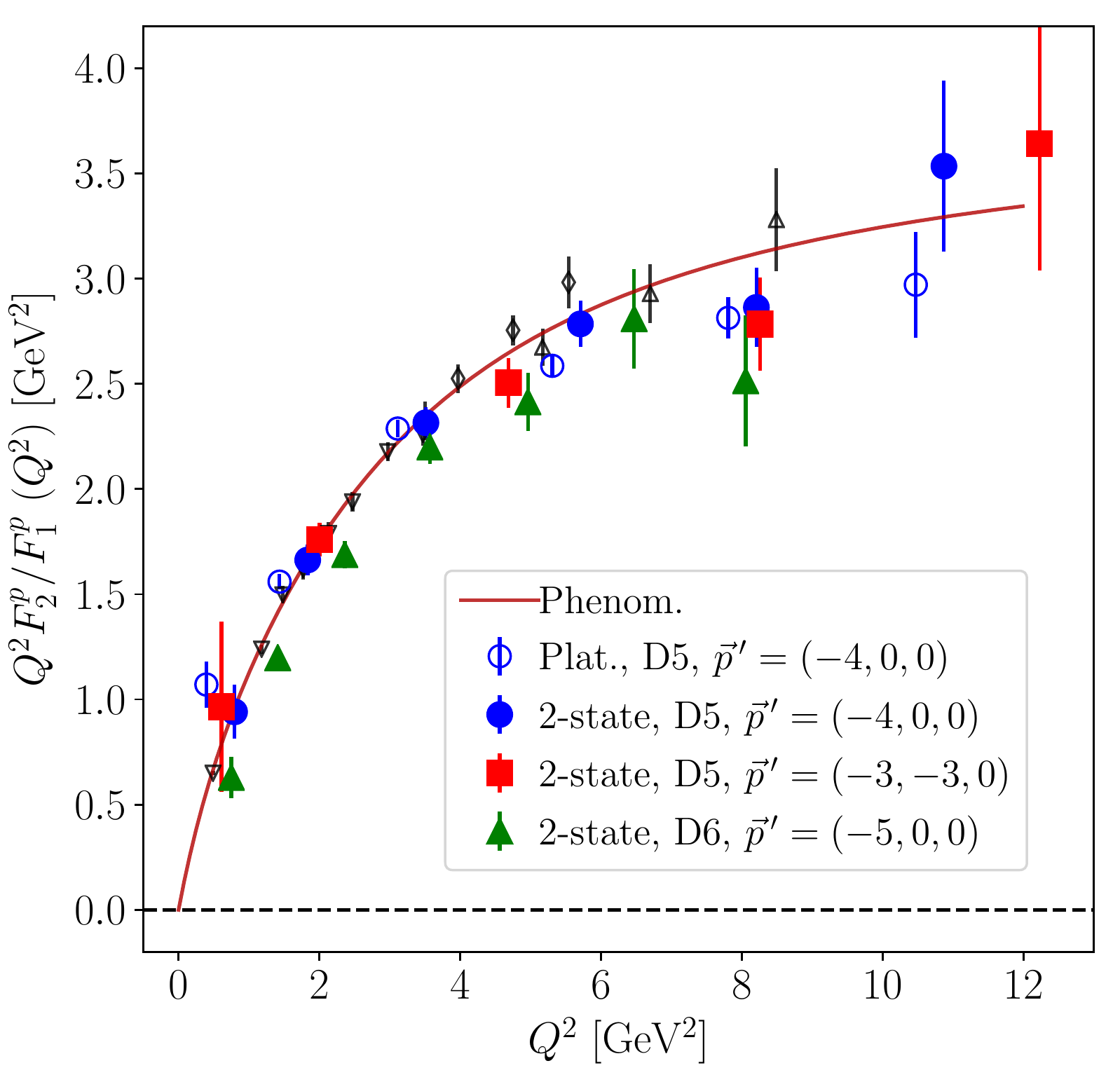}
\end{minipage}\hfill
\begin{minipage}{0.33\textwidth}
\includegraphics[width=\textwidth]{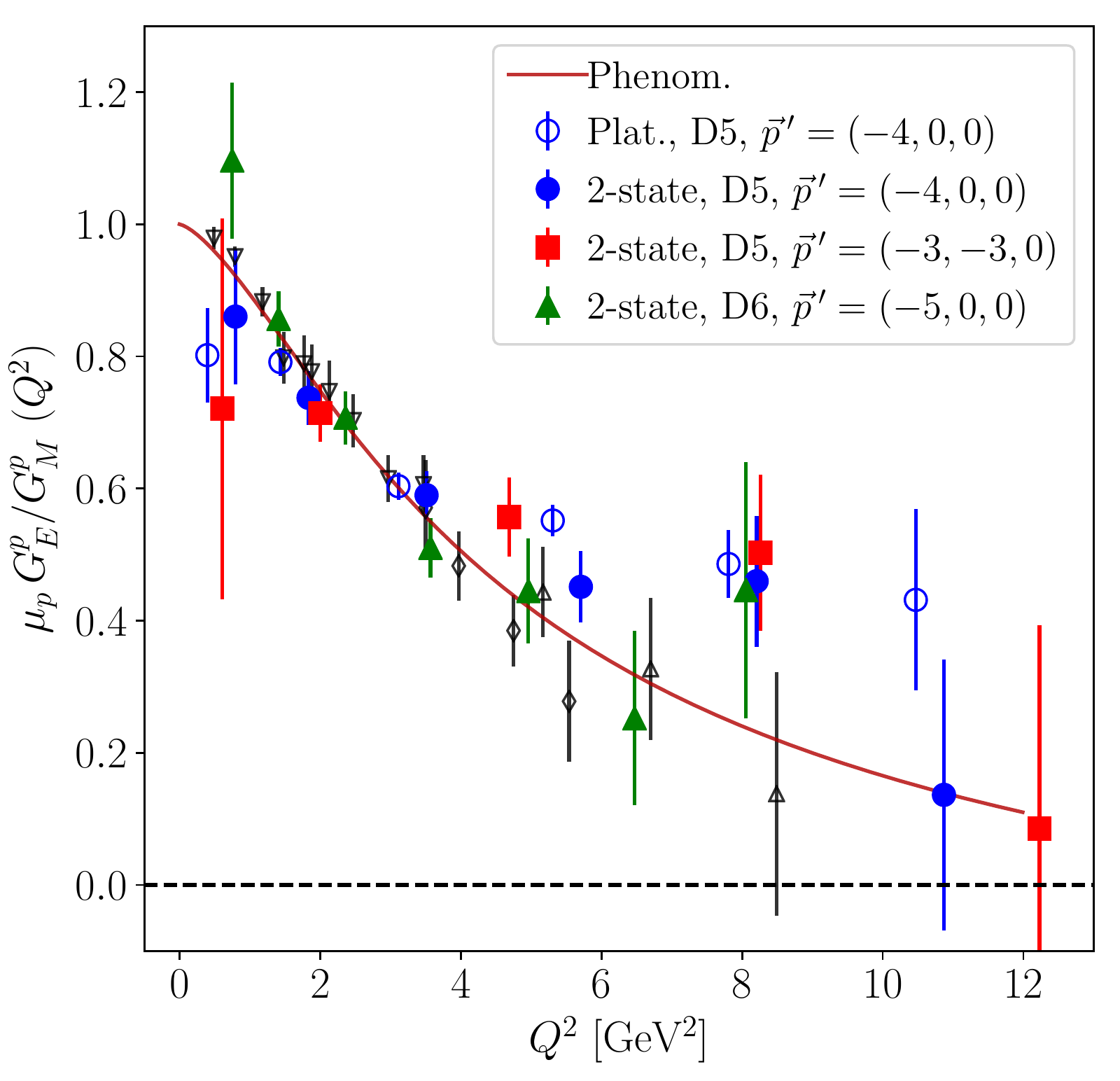}
\end{minipage}
\begin{minipage}{0.33\textwidth}
\includegraphics[width=1.03\textwidth]{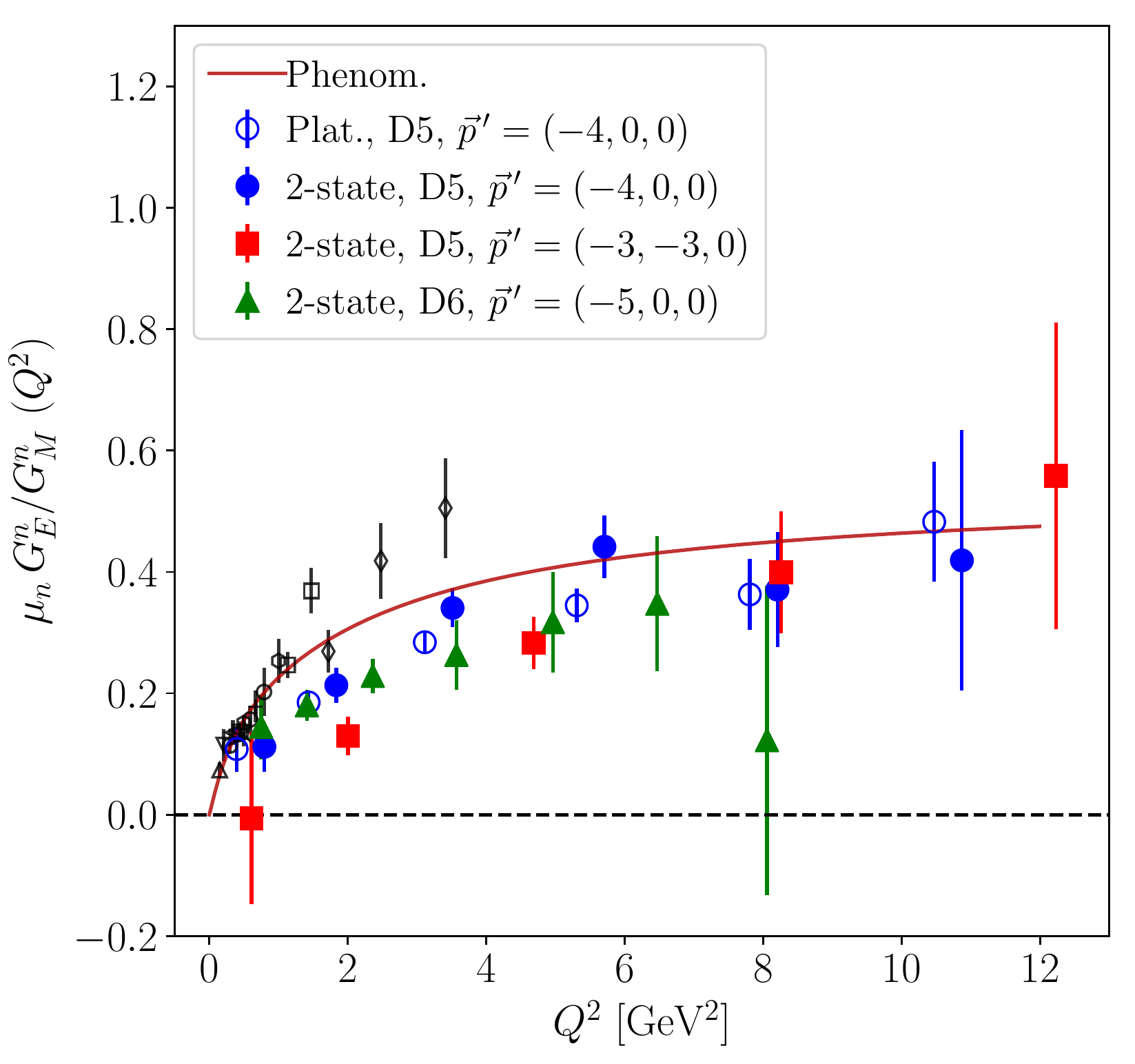}
\end{minipage}
\begin{minipage}{0.33\textwidth}
\includegraphics[width=0.97\textwidth]{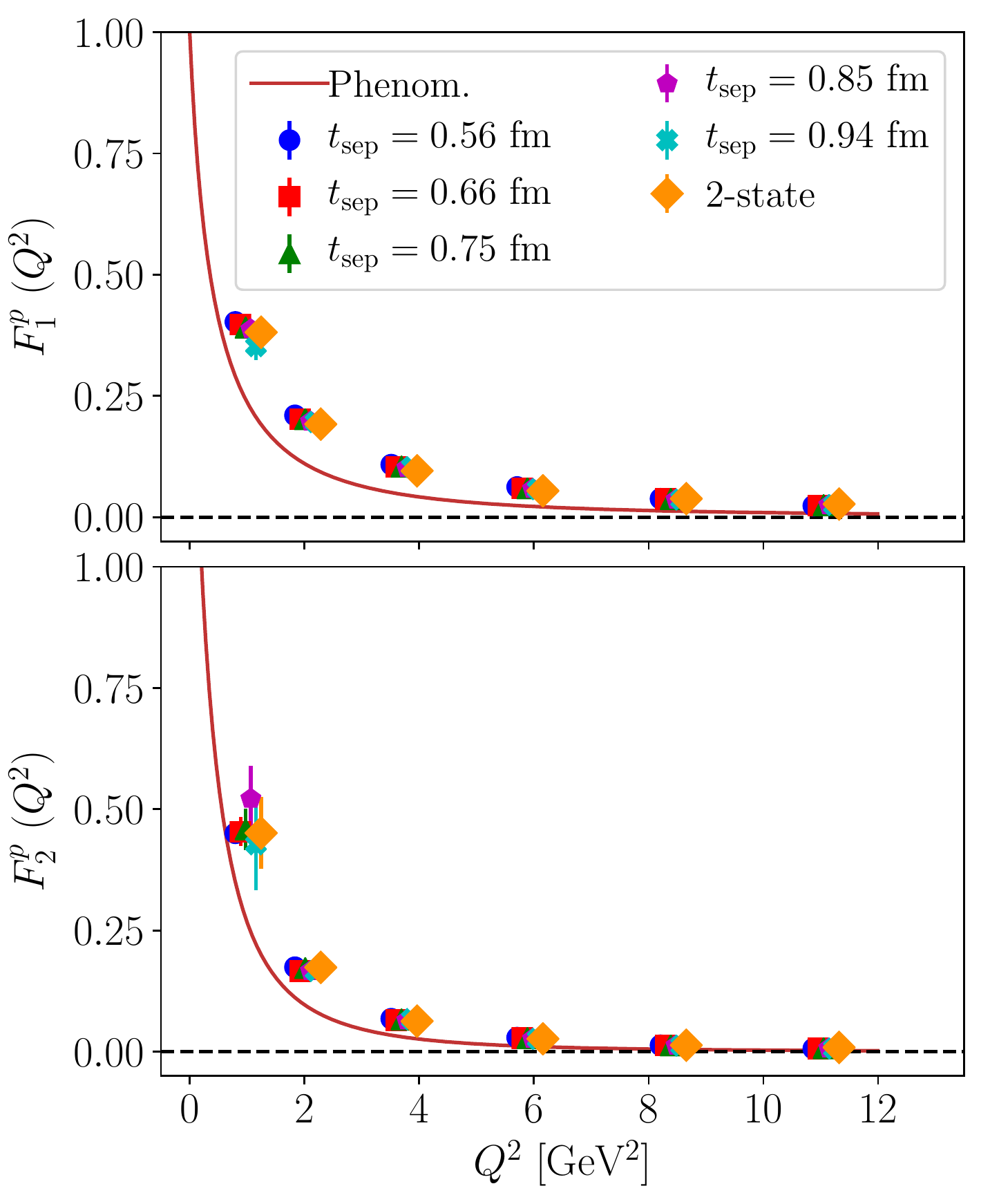}
\end{minipage}\hfill
\begin{minipage}{0.33\textwidth}
\includegraphics[width=0.95\textwidth]{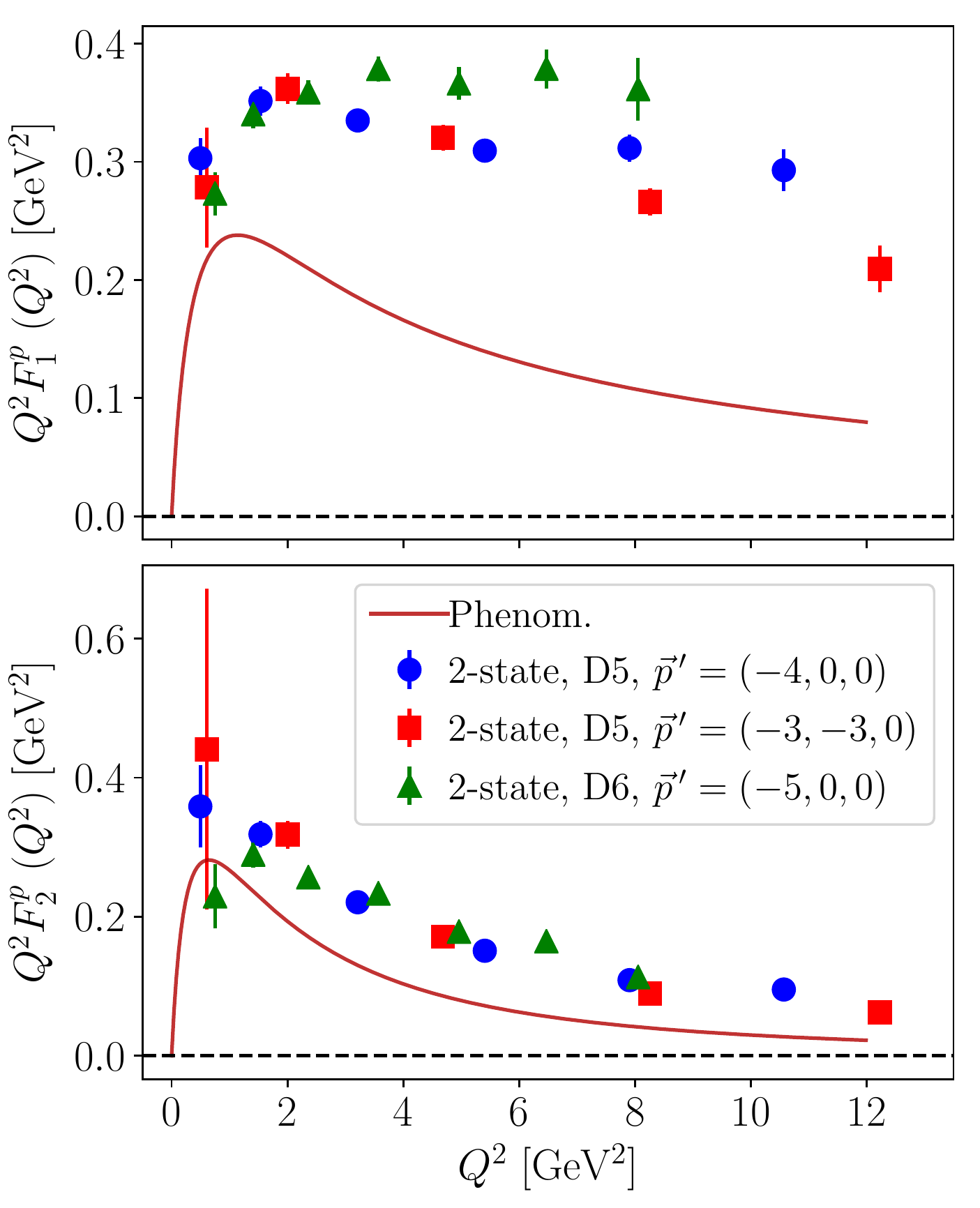}
\end{minipage}
\begin{minipage}{0.33\textwidth}
\includegraphics[width=\textwidth]{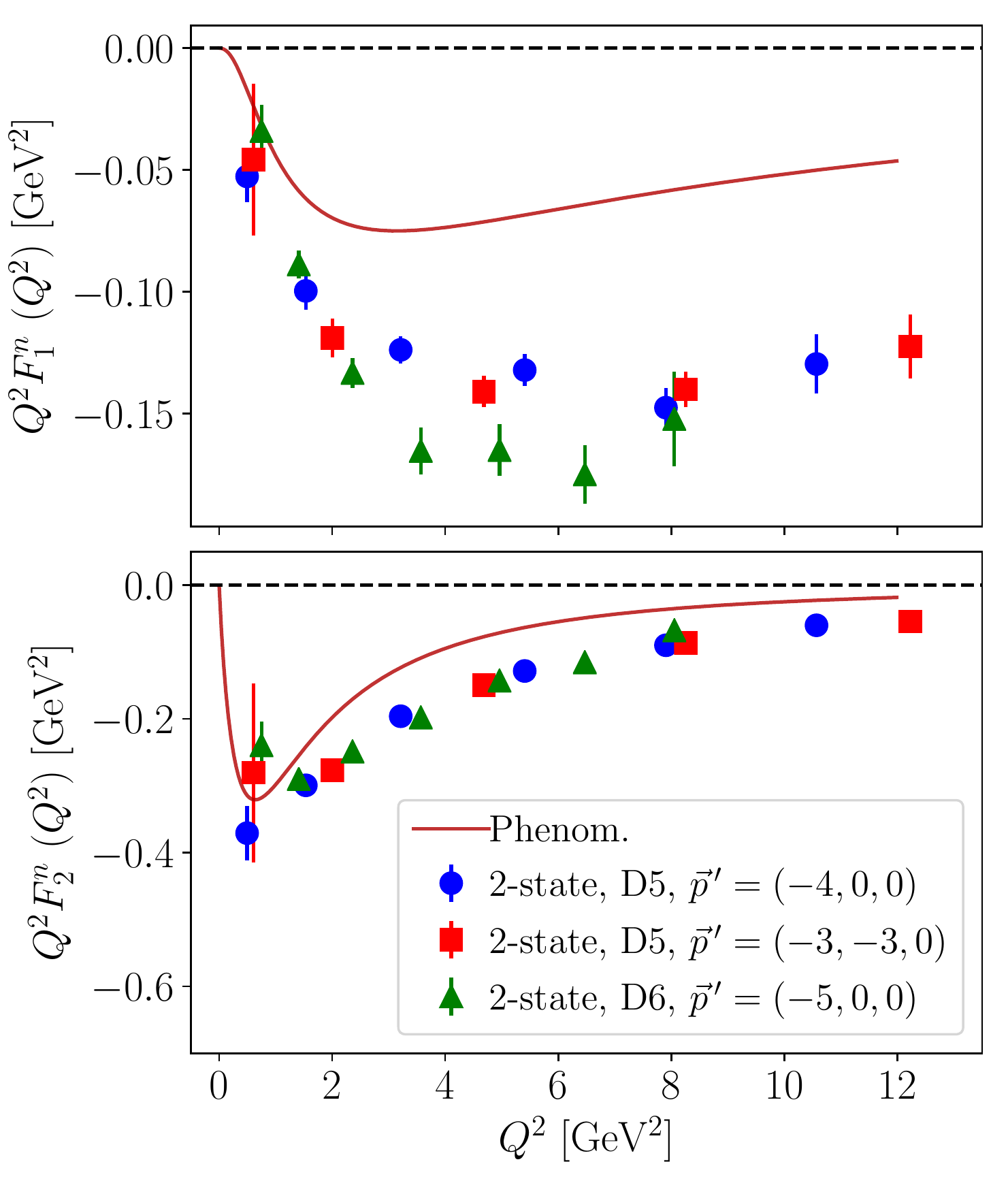}
\end{minipage}
\caption{{\bf Top}: The $Q^2$-dependence of the ratios: $Q^2 F_2/F_1$ for the proton (left), $\mu_p G_E/G_M$ for the proton (center) and $\mu_n G_E/G_M$ for the neutron (right). The experimental data are shown with the open black lower~\cite{Jones:1999rz} and upper~\cite{Puckett:2010ac} triangles, diamonds~\cite{Gayou:2001qd} and squares~\cite{Madey:2003av}. {\bf Bottom}: The $Q^2$-dependence of the proton $F_1$ and $F_2$ (left), and the two-state fit results from all ensembles and sink momenta for the proton (center) and neutron (right) $F_1$ and $F_2$, scaled with $Q^2$. The phenomenological curves are taken from Ref.~\cite{Alberico:2008sz}.}
\label{fig:ffPlots}
\vspace*{-0.25cm}
\end{figure}

We begin the presentation of our results by showing the $Q^2$-dependence of the proton form factor ratio $F_2^p/F_1^p$ scaled with $Q^2$, as well as the ratios $G_E/G_M$ for the proton and neutron, scaled with the corresponding magnetic moment $\mu_p$ and $\mu_n$ on the top panel of~\fig{fig:ffPlots}. In the plots we include the two-state fits from both the D5 and D6 ensembles and for all the sink momenta analyzed. We also include the plateau fits from the D5 ensemble at $t_{\rm sep}=0.94$~fm and sink momentum $\vec{p}'=\frac{2\pi}{L}(-4,0,0)$. As a general remark, we note that in these three plots, both the D5 and D6 ensembles for all values of sink momentum yield consistent results within our statistical uncertainty.

Concerning the ratio $Q^2 F_2^p/F_1^p$, we find very good agreement with the experimental data~\cite{Jones:1999rz,Gayou:2001qd,Puckett:2010ac,Madey:2003av} as well as with the recent phenomenological parametrization from Alberico et al.~\cite{Alberico:2008sz} across our range of $Q^2$ values. Given the agreement with experiment, our lattice data also support the pQCD scaling $Q^2 F_2^p/F_1^p \sim {\rm log}^2[Q^2/\Lambda^2]$ suggested by Belitsky et al.~\cite{Belitsky:2002kj} where $\Lambda$ is a non-perturbative mass scale, rather than the originally predicted $F_2/F_1\sim 1/Q^2$ behavior by Brodsky and Farrar~\cite{Brodsky:1974vy}. This logarithmic scaling hints that quarks carry sizable orbital angular momentum, and it plays an important dynamical role in the $Q^2$-evolution of the proton form factors, forming the basis for the prediction~\cite{Belitsky:2002kj}. Therefore, it is crucial to improve our statistical and systematic uncertainties, and potentially include results at $Q^2>12$~GeV$^2$ within this project, in order to reliably clarify the $F_2/F_1$ behavior from the lattice.

\vspace{-0.2cm}

Concerning the proton $\mu_p G_E^p/G_M^p$, our data are consistent with the experimental data and phenomenology, though for $Q^2 > 6~$GeV$^2$ our uncertainties become too large to extract a safe conclusion. Nevertheless, it is clear that our results support a smoother approach of the ratio towards zero, as predicted by two vector-meson dominance (VMD) models~\cite{Lomon:2001ga}, whereas other theoretical calculations, e.g. Ref.~\cite{Belitsky:2002kj,Miller:2002qb}, predict $G_E^p/G_M^p$ to cross zero at $Q^2\approx 7$~GeV$^2$.

Focusing on the neutron $\mu_n G_E^n/G_M^n$, where experimental data exist only up to $Q^2\simeq 3.4$~GeV$^2$, we observe that our results consistently underestimate the experiment and the phenomenological prediction by Alberico et al.~\cite{Alberico:2008sz}. However, the qualitative behavior is very similar. This difference might be due to disconnected contributions, which we neglect in this calculation. We anticipate that with a full investigation of systematics and excited state effects this discrepancy will be reduced.

We continue our discussion by showing representative results on the proton and neutron individual form factors $F_1$ and $F_2$ on the bottom panel of~\fig{fig:ffPlots}. On the bottom left, we plot the $Q^2$-dependence for $F_1^p$ and $F_2^p$, showing all five $t_{\rm sep}$ and the two-state fit, from the D5 ensemble at sink momentum $\vec{p}'=\frac{2\pi}{L}(-4,0,0)$. As indicated by the data, results with different $t_{\rm sep}$ are compatible, and are also in agreement with the two-state fit. However, these results overestimate phenomenology across our range of $Q^2$ values. In order to display this discrepancy, we plot $F_1$ and $F_2$ for the proton and neutron, scaled with $Q^2$, on the bottom center and right panels of~\fig{fig:ffPlots}, respectively. In these plots we show the results from the two-state fits, including all ensembles and sink momenta. Concerning the $F_1$ form factor for both proton and neutron, the lattice results from D5 with $\vec{p}'=\frac{2\pi}{L}(-3,-3,0)$ (red squares) follow the qualitative behavior of the phenomenological description better than the other two cases, which appear to not follow the curve, especially for $Q^2>4$~GeV$^2$. On the other hand, our results for the proton and neutron $F_2$ from all ensembles and sink momenta follow the qualitative behavior of phenomenology quite well, while still overestimating it by absolute value. These discrepancies observed for the individual form factors can be attributed to various systematic effects discussed in the next section. However, it is still remarkable that the discrepancies seem to vanish in the ratios of form factors.

\vspace{-0.2cm}


\section{Discussion, Conclusions, and Outlook}\label{sec:concl}

In this lattice calculation we obtain high statistics results for the nucleon EM form factors for momentum transfer up to about $Q^2=12$~GeV$^2$, from two Wilson-clover gauge ensembles at pion masses $m_\pi \approx 170$~MeV and $m_\pi \approx 280$~MeV. In the analysis presented here, only statistical errors are considered. As discussed in~\sec{sec:results}, we find very good agreement with experiment and phenomenology for the proton and neutron ratios $F_2/F_1$ and $G_E/G_M$. However, we observe discrepancies of multiple $\sigma$ when looking at the $Q^2$-dependence of the individual form factors.  

We first note that pion mass and volume effects appear to be small, given that all of our results show a broad agreement. We also do not observe any significant dependence on the source-sink separation, making large excited state contributions unlikely. Moreover, we have studied possible contamination of our nucleon correlators with negative-parity states using Parity-Expanded Variational analysis (PEVA)~\cite{Menadue:2013kfi} and found that these effects are negligible within our statistical accuracy (this will be presented in a separate publication). 
 
Another systematic effect is the breaking of rotational symmetry in the source construction due to momentum smearing. Comparing our current results from the D5 ensemble with two orientations of nucleon sink momenta, we do not observe any significant discrepancies. The only exception is the case of the proton and neutron form factor $F^{p,n}_1$, where we observe that results obtained with diagonal boosting momentum are in better qualitative agreement with the phenomenological fits. A separate study of rotational symmetry breaking in the nucleon propagator is underway.

The relation between the lattice wave vector $\vec{\kappa}=\frac{2\pi}{L}\vec{n}$ ($\vec{n}=\frac{1}{a}[-L/2,L/2)$) and the nucleon momentum $\vec{p}$ is also a potential systematic issue. In our study we adopt the commonly accepted convention that the momentum $\vec{p}$ entering the nucleon polarization matrix $-i\slashed{p}+m$ is linear in $\vec{\kappa}$. However, this relation holds only up to $\mathcal{O}(a^2\kappa^3)$, and may be subject to discretization effects, especially at large $\vec n$ \footnote{One can compare that to  the ``momentum'' from nearest-neighbor difference,  $(\vec p)_i=\frac1a\sin(a\kappa_i)\approx\kappa_i(1-\frac16(a\kappa_i)^2)$.}. An investigation on this effect is currently ongoing.


A further pending calculation is the $\mathcal{O}(a)$ improvement of the vector current, which is expected to provide a correction of about $5\%-10\%$ in our lattice results. This procedure is of paramount importance in order to assure that the $\mathcal{O}(a)$ discretization effects of the Wilson-clover fermion action are removed. Finally, a future calculation that would complete this project is the inclusion of contributions from disconnected diagrams, which require special techniques and large amounts of computational power to overcome bad signal-to-noise ratio, especially at high momentum~\cite{Syritsyn:2017jrc}.

Given the quality of our results, this calculation is a strong indication that accessing large $Q^2$ from the lattice is feasible, provided that proper techniques are employed in order to control the statistical errors. With a careful and complete investigation of all systematic uncertainties, a lattice calculation of the nucleon EM form factors at high $Q^2$ can provide robust predictions and input to current and future related experiments.

{\bf Acknowledgements:} We are grateful to Kostas Orginos for providing the gauge ensembles, generated using resources provided by XSEDE (supported by National Science Foundation Grant No. ACI-1053575). 
S. M. is supported by the U.S. Department of Energy (DoE), Office of Science, Office of High Energy Physics under Award Number DE-SC0009913.
S. S. and S. M. also acknowledge support by the RHIC Physics Fellow Program of the RIKEN BNL Research Center.
M. E., J. N, and A. P. are supported by the U.S. DoE, Office of Science, Office of Nuclear Physics through grants numbered DE-FG02- 96ER40965, DE-SC-0011090 and DE-FC02-06ER41444 respectively.
Computations were performed in facilities of the USQCD Collaboration, using the Qlua and QUDA software suites.

\bibliographystyle{./apsrev.bst}
\tiny{\bibliography{refs}}

\end{document}